# An extensive impurity-scattering study on the pairing symmetry of monolayer FeSe films on SrTiO$_3$

Chong Liu[1], Jiahao Mao[2], Hao Ding[1], Rui Wu[1], Chenjia Tang[1], Fangsen Li[1], Ke He[1,3], Wei Li[1,3], Can-Li Song[1,3], Xu-Cun Ma[1,3], Zheng Liu[2†], Lili Wang[1,3†], and Qi-Kun Xue[1,3†]

[1]*State Key Laboratory of Low-Dimensional Quantum Physics, Department of Physics, Tsinghua University, Beijing 100084, People's Republic of China*
[2]*Institute for Advanced Study, Tsinghua University, Beijing 100084, People's Republic of China*
[3]*Collaborative Innovation Center of Quantum Matter, Beijing 100084, People's Republic of China*

†Correspondence to: zheng-liu@mail.tsinghua.edu.cn, liliwang@mail.tsinghua.edu.cn; qkxue@mail.tsinghua.edu.cn

Determination of the pairing symmetry in monolayer FeSe films on SrTiO$_3$ is a requisite for understanding the high superconducting transition temperature in this system, which has attracted intense theoretical and experimental studies but remains controversial. Here, by introducing several types of point defects in FeSe monolayer films, we conduct a systematic investigation on the impurity-induced electronic states by spatially resolved scanning tunneling spectroscopy. Ranging from surface adsorption, chemical substitution to intrinsic structural modification, these defects generate a variety of scattering strength, which renders new insights on the pairing symmetry.

**Introduction**

Interface high temperature superconductivity in monolayer FeSe films grown on SrTiO$_3$ substrates has appealed great interest due to its intriguing interfacial effects and highest superconducting transition temperature ($T_c$) among iron-based superconductors. It exhibits a characteristic double-full-gap with magnitude of $\Delta_1 \sim 10$ meV and $\Delta_2 \sim 15\text{-}20$ meV [1-3] which closes over 65 K [4, 5]. Regardless of the very high transition temperature, its Fermi surface (FS) consists of only electron-like pockets centered around the Brillouin zone (BZ) corners and there are no hole pockets at the BZ center that are typical in bulk iron-based superconductors, as a result of heavy interface electron doping [4-6]. The absence of hole pockets at the BZ center directly challenges the previously perceived $s_\pm$ pairing scenario that relies on strong spin fluctuation induced by the repulsive interband interaction between the hole bands around BZ center and the electron bands around BZ corner [7, 8].

Determination of the pairing symmetry in monolayer FeSe on SrTiO$_3$ is a requisite for understanding the mechanism of high temperature superconductivity. Theoretical studies have proposed various pairing symmetries for such systems with only electron pockets, such as plain *s*-wave pairing [9-11], 'quasi-nodeless' *d*-wave pairing [12, 13], several new types of $s_\pm$ pairing that involve the 'folding' of the Brillouin zone and band hybridization [14], orbital dependent pairing [15], or mixing of the even- and odd-parity pairing [16]. Very recently, an anisotropic $s_{++}$ -wave state due to inter-electron-pocket pairing was proposed for heavily electron-doped FeSe [17].

An experimental pathway to obtain information on pairing symmetry is to study atomic-scale impurity scattering by using scanning tunneling microscopy/spectroscopy (STM/STS) [18, 20, 21]. It is well known in theory that only magnetic impurities are pair breakers in the plain s-wave superconductors [18], whereas if the pairing order parameter is sign-changing both magnetic and nonmagnetic impurities can induce in-gap states and suppress superconductivity [19]. By introducing surface impurities on the monolayer FeSe surface, earlier STM/STS investigations indicate that the superconductivity is locally suppressed at magnetic impurities (such as Cr and Mn) but remains stubborn at non-magnetic impurities (such as Zn, Ag and K) [20]. Combining the quasi-particle interference patterns and their response to magnetic field, the authors concluded a plain *s*-wave pairing symmetry in monolayer FeSe on SrTiO$_3$ [20].

In contrast, recently, Zn substitutional impurities at Fe-site in Li$_{1-x}$Fe$_x$OH intercalated FeSe compound, i.e. (Li$_{1-x}$Fe$_x$)OHFe$_{1-y}$Zn$_y$Se, which has similar FS geometry and a $T_c$ of 40 K, has been investigated [21]. With the observed in-gap bound states on substitutional Zn impurities, the authors concluded a sign-changing order parameter and pointed out that the spin fluctuation plays a key role in electron pairing [21]. It is also worth mentioning that recent high-resolution angle-resolved photoemission spectroscopy (ARPES) measurements on monolayer FeSe revealed two overlapped ellipse-like electron pockets and a peculiar nodeless anisotropic superconducting gap in each of the electron pockets [22]. In this study, the occurrence of four gap minima at the intersections of electron pockets is interpreted as a result of weak inter-pocket hybridation on a nodeless d-wave gap or a competition between intra- and interorbital pairing [22]. These new results cast doubt on the s-wave pairing symmetry.

Although nonmagnetic impurities in principle serve as a sensitive probe to distinguish sign-preserving pairing from sign-changing pairing, the practical situation is not that simple. A problem is that it is hard to prove whether the impurities under investigation are magnetic or not - sometimes a nominally nonmagnetic impurity may pin local magnetism of the parent material [23], and sometimes a commonly-perceived magnetic element appear to lose its local moment when embedded in the parent material [24]. Another problem is that, when the impurity-induced scattering is weak, the bound state lies close to the gap edge [18]. It then becomes tricky to resolve the bound-state peak from the superconducting coherence peak. Consequently, experiment based on a single impurity type could be ambiguous, and often lead to controversial conclusions, *e.g.* sign preserving supported by impurity effect induced by surface Zn atom in Ref. 20 *vs.* sign changing by Zn substitution in Ref. 21 as addressed earlier.

The motivation of the present work is to systematically investigate impurity scattering effects by introducing a series of different impurities. The impurities include surface K adatoms, substitutional Co ($Co_{Fe}$) and Cu ($Cu_{Fe}$) impurities at Fe sites of FeSe films. We have also checked native defects in FeSe, such as Fe vacancy, Se vacancy and $Se_{Fe}$ antisite defect. We further deposited Sr on the $SrTiO_3$ substrates to introduce Sr impurities. These impurities are expected to generate a variety of scattering strength on the electronic states near impurity sites, which guarantee systematically examination of the pairing symmetry using STM/STS.

**Experiments**

The Nb-doped (0.05 wt. % ) $SrTiO_3$(001) substrates were annealed at 1150°C for 15 min before FeSe deposition to obtain double-$TiO_2$ terminated surface [3]. The monolayer FeSe films were grown at a substrate temperature of 400 °C with a flux rate of 0.25 monolayer (ML) per minute as previously reported [1]. To introduce $Co_{Fe}$ ($Cu_{Fe}$) impurities, we co-deposited Co (Cu) and Se onto monolayer FeSe films at 400 °C with a flux rate of 0.01 ML per minute and then annealed the films up to 480 °C to remove extra Se atoms and enhance mixing of the successively deposited FeSe and CoSe (CuSe). After such treatment, spatially uniform $Fe_{1-x}Co_xSe$ ($Fe_{1-x}Cu_xSe$) films formed. Surface K atoms were deposited on monolayer FeSe films cooled with liquid nitrogen. In the case of the Sr-treatment of $SrTiO_3$ substrates, we deposited about one monolayer Sr on double-$TiO_2$ terminated $SrTiO_3$ at a substrate temperature of 950 °C and a flux rate of 0.10 ML per minute. In all STM/STS measurements, a polycrystalline PtIr tip was used. Tunneling spectra were acquired at sample temperature of 5 K using a standard lock-in technique with a bias modulation of 0.4 mV at 937 Hz, unless otherwise specified.

**Results**

A. Intrinsic impurities in monolayer FeSe films

Figure 1 presents the STM characterization of monolayer FeSe films on $SrTiO_3$(001). The STM images with atomic resolution in Figs. 1(a) and (b) show brighter dumbbell-like pairs in ordered Se-terminated FeSe(001) surface with lattice constant $a_0 \sim 0.39$ nm. The dumbbell-like pairs consist of two bright lobes on adjacent top-layer Se sites with the center located at the Fe site beneath the top Se layer, which has been interpreted as perturbation at the Fe site that affects the orbitals on neighboring Se atoms [25]. There are several candidates to induce perturbation at the Fe-site, such as Fe-site vacancies, $Se_{Fe}$ antisite defects, Fe-site substitutions, as will be shown below.

The d$I$/d$V$ spectra of the monolayer FeSe films shown here have a double-gap feature with inner coherence peaks located at ±10 meV and outer ones at ± 16 meV, as indicated by dashed lines in the spectra shown in Fig. 1(c) and 1(d). The result is consistent with previous results [1] within experimental uncertainty. At the dumbbell pair sites shown in Fig. 1(a), the coherence peaks at ± 10 meV are enhanced, especially the one below Fermi level is more strongly enhanced than the one above Fermi level, while the zero bias conductance (ZBC) remains at zero. This impurity effect is visible within one lattice distance $a_0$ and disappears at a distance of 1.5 $a_0$ (Fig. 1(c)). In contrast, at dumbbell pair sites shown in Fig. 1(b), an in-gap bound state occurs at 3 meV below Fermi level, which lifts ZBC to a finite value, as shown in Fig. 1(d). Again, the impurity effect is visible within one lattice distance $a_0$ and disappears at a distance of 1.5 $a_0$. The sharp contrast in spectral features suggests that the two dumbbell pairs could originate from distinct defects. Indeed, they also exhibit different behavior during annealing. The defects with the spectral feature shown in Fig. (c) are usually observed under Se-rich condition and disappear gradually after annealing at 450-480 °C, indicating that it corresponds to a Fe vacancy [26]. Previously they were assigned as extra Se atoms, in view of the comparative stoichiometry [3, 27]. On the other hand, the defects with the spectral feature shown in Fig. 1(d) are stubborn and can't be removed by annealing at 450-480 °C, which attests their origin as $Se_{Fe}$ antisite defects.

We regard the third type of intrinsic defect shown in Fig. 1(e) as Se vacancy, for it shows the same morphology as Se vacancy in multilayer FeSe films [27], i.e. it corresponds to a missing Se atom with its four nearest neighbor (NN) Se atoms exhibiting enhanced apparent height in comparison with ordered Se atoms in the background. Within $4a_0$ around the missing Se atom, the in-gap bound states occur at -4 meV. We observed much stronger in-gap bound state on the NN Se atoms than on the center site (Fig. 1(f)), which is consistent with the contrast in morphology shown in Fig. 1 (e).

B. Artificial Fe-site impurities

Figures 2(a)-(c) present the topographic images of $Fe_{1-x}Co_xSe$ films with (a) x = 0.002, (b) x = 0.02 and (c) x = 0.1, where the values of x were estimated from the coverage ratio between the deposited Co and Fe [28]. In all of these images, one can see brighter dumbbell-like pairs along either [100] or [010] directions in the background of ordered square lattices, whose density increases with increasing Co coverage. The contrast is enhanced when many dumbbell-like pairs gather (marked by arrows in Fig. 2(b) and 2(c)). We compared the density of dumbbell-like pairs with the Co/Fe ratio and found that there is a consistency between the number of dumbbell-like pairs and the Co coverage [28]. Thus we argue that each dumbbell-like pair observed here corresponds to an individual $Co_{Fe}$ substitution at Fe-site, except for some of them that are $Se_{Fe}$ antisite-defects as identified from $dI/dV$ spectra, which will be addressed later on.

Depicted in Fig. 2(d) are the typical tunneling spectra of monolayer $Fe_{1-x}Co_xSe$ films, which were taken at the locations away from the dumbbell-like pairs. For comparison, we also show a spectrum of pure monolayer FeSe films taken on the same sample but before depositing CoSe. Clearly, the coherence peaks are nearly unaltered at x = 0.002. With increasing Co component, the coherence peaks becomebroadened. The two pairs of coherence peaks are almost unresolvable at x = 0.02, while the ZBC stays close to 0. At x = 0.1, no superconducting gap can be observed at all, indicating that the superconductivity is destroyed completely.

In order to have a detailed description on the local density of states induced by the $Co_{Fe}$ substitution, we did the spatial-resolved STS measurements with spacing of lattice constant $a_0$, as marked by the arrow in Fig. 3(a), in the sample with x = 0.002. The raw $dI/dV$ spectra taken in a large (small) bias range from -500 mV to 500 mV (from -40 mV to 40 mV) are shown in Fig. 3(b) (Fig. 3(c)). As shown in Fig. 3(b), near the $Co_{Fe}$ substitution impurity, the kink in the valence band at -250 meV shifts downward to -300 meV, which is similar to the observation on K-doped FeSe [2], indicative of local electron doping induced by Co substitution of Fe atoms. Moreover, in a spatial range of $a_0$ around $Co_{Fe}$ substitution, resonance peaks appear at approximately 9 meV, close to the inner-gap edge, while the ZBC remains at zero and the coherence peak at around -15 mV seems to be suppressed (Fig. 3(c)). Displayed in Fig. 3(d) and 3(e) are the d$I$/d$V$ mapping at 10 meV and -15 meV, respectively. One can see a systematic evolution of the local density of states around the $Co_{Fe}$ substitution with a spatial scale of 0.6 - 1.0 nm, same as the size of the dumbbell shape shown by the topographic image in Fig. 3(a).

Similar to the $Co_{Fe}$ substitution impurity, the $Cu_{Fe}$ substitution also appears as dumbbell-like pair, as shown in Fig. 3(f). However, it generates a pronounced in-gap bound state at + 4.5 mV (Fig. 3(g)). As a brief summary on Fe-site impurities, Fe vacancy and $Co_{Fe}$ substitution induce subtle change near inner gap edge, whereas $Cu_{Fe}$ substitution and $Se_{Fe}$ antisite defect induce in-gap bound states above and below Fermi level, respectively.

C. Surface K-atom and Sr-related defects

Figure 4 (a) displays the morphology of the monolayer FeSe films with 0.01 monolayer K adsorption. Potassium atoms distribute randomly and occupy the hollow sites of 1×1 lattice of topmost Se layer [2]. The K atoms appear as single bright dots and contribute somehow charge transfer to monolayer FeSe films [29]. From the measured local density of states on surface K-atom, the superconductivity is basically undisturbed since most of our observation is consistent with previous results [20]: the coherence peaks remain almost unchanged and there is no in-gap bound state. Only occasionally, we observed change in coherence peaks. An example is shown in Fig. 4(b). From the series of d$I$/d$V$ spectra taken across a K-atom, we can see that the coherence peaks at -10 meV are enhanced while those at +10 meV are suppressed. It is worth pointing out the possibility of intrinsic FeSe impurities below/around this K atom. Surface K-atom appears much brighter than the background atoms, which can cover the morphology feature, *i.e.* the dumbbell-like pair for intrinsic Fe vacancy as shown in Fig.1a. Moreover, their spectra feature, enhanced coherence peaks at negative bias but suppressed at positive bias, is also similar.

Displayed in Figs. 4(c) and 4(d) are the topography images of the Sr-treated SrTiO$_3$ substrate and the monolayer FeSe grown on it, respectively. Instead of 2 × 2 reconstruction on TiO$_2$-terminated SrTiO$_3$ surface, the Sr-treated surface exhibits unreconstructed 1×1 lattice with a lattice constant of 0.39 nm, which could be a result of Sr intercalation into the surface double-TiO$_2$ layer [30]. In addition, two types of defects with four-fold symmetry appear as quatrefoils with the center at the hollow site of the topmost 1×1 surface (Fig. 4(c)). After the monolayer FeSe was deposited, a new type of defects with four-fold symmetry occurs. As shown in Fig. 4(d), this kind of defect appears as quatrefoil centered at surface Se atom site with the center and its four NN Se atoms showing higher apparent height than the Se atoms in the background. Its morphology is similar to Se vacancy shown in Fig. 1(e) with the same four-fold symmetry and bright lobes at NN sites, except that its center appears bright instead of dark. This type of defects is exclusively observed in monolayer FeSe on the Sr-treated SrTiO$_3$ substrate but not in 2-UC FeSe films or monolayer FeSe on other substrates, which attests its correlation with Sr-treatment. Similar to Se vacancy again, such defects induce strong resonance peaks below the Fermi level and remarkable increase in ZBC in turn (Fig. 4(e)). Here the in-gap bound state occurs at a little larger energy of 5 meV below the Fermi level, compared with the in-gap bound states at 4 meV at Se vacancy (Fig. 1(f)).

**Discussion**

The impurity state in a superconductor with given pairing symmetry has been extensively studied [18]. For iron-based superconductors, the additional complexities are the multi-orbital nature. Consequently, the impurity-induced scattering should be orbital-dependent, and the involved orbitals, the band dispersion and gap anisotropy were found to be relevant [31-33]. In principle, a proper parameterization of all these quantities should be able to reproduce the realistic local density of states, which renders a direct comparison with the d$I$/d$V$ spectrum. In practice however, this remains to be a highly challenging task, especially for such a strongly-correlated interfacial system. Nevertheless, some qualitative differences between the sign-preserving and sign-changing scenarios are robust against the microscopic details, such as the existence/absence of the bound-state solution and the number of solutions. Accordingly, we discuss our experimental data from these two aspects.

(a) *The existence/absence of the bound-state solution.* It is well established, commonly known as Anderson's theorem [34], that the bound-state solution strongly relies on the sign structure of $\Delta(k)$: if $\Delta(k)$ is sign-preserving, i.e. for s-wave superconductors, only magnetic scattering are pair breakers, whereas if $\Delta(k)$ is sign-changing, potential scattering alone can induce in-gap states and suppress superconductivity [18]. Within the isotropic approximation, the dimensionless scattering strength $\tilde{U}$ (nonmagnetic) and $\tilde{J}$ (magnetic) fully dictate the bound-state energy. The analytical formulas are derived in details in the Appendix.

In our experiment, there are three defect types that do not exhibit well-defined bound-state resonances, including surface K-atom, Co$_{Fe}$ substitution and intrinsic Fe vacancy. Modulation right at the gap edge can be observed, but it is difficult to tell whether a new bound-state peak emerges in addition to the original superconducting coherence peak. In contrast, for Cu$_{Fe}$ substitution, Se$_{Fe}$ antisite defect, Se vacancy, and Sr-related defect, clear bound-state resonances are resolved.

If we adopt the sign-preserving scenario [c.f. Eq. (A3)], it requires that the latter four defects generate noticeable magnetic scattering, *i.e.* $\tilde{J} \neq 0$, while the former three do not, *i.e.* $\tilde{J} \sim 0$. One concern is that similar Cu$_{Fe}$ impurities in bulk NaFeAs have been reported to be nonmagnetic based on macroscopic magnetization measurement [35]. However, for a monolayer, the determination of magnetic impurities is much more difficult. Indeed, even for the pristine FeSe monolayer, the magnetic state of the iron plane is still in controversy.

If we adopt the sign-changing scenario [c.f. Eq. (A4)], the latter four defects could be either magnetic or nonmagnetic scatterers, but the former three defects must not generate any noticeable scattering, *i.e.* $\tilde{U} \sim \tilde{J} \sim 0$, or the scattering potential has a special form that almost does not frustrate the phase structure of the pairing function. From this aspect, constraint on the sign-changing scenario is more rigid. Especially, the Fe vacancy defect is intuitively a strong scatterer ($\tilde{U} \neq 0$). Clear electron density modulation was observed around the Fe vacancy in our previous work [27].

(b) *The number of solutions.* For the sign-changing case, the bound state induced by a nonmagnetic impurity is doubly degenerate owing to the time-reversal symmetry. When the magnetic scattering is also present, the double degeneracy can be removed, which typically manifests in two pairs of resonance peaks in STS [c.f. Eq. (A4)]. This feature was first predicted in theory for the *d*-wave cuprates [36] and later observed in experiment [37]. In contrast, in the sign-preserving case, the bound state is typically a spin-polarized single level arising from the magnetic scattering

For $Cu_{Fe}$ substitution, $Se_{Fe}$ antisite defect, Se vacancy, and Sr-related defect, which show clear bound-state resonances, only one in-gap peak can be resolved. This observation natural fits in the sign-preserving scenario [c.f. Eq. (A3)]. Note that the asymmetric d$I$/d$V$ intensity at the positive and negative resonance energy is a general property of superconductors as seen in both s- and d-wave systems [18]. If we adopt the sign-changing scenario, the possibility is that the two pairs of peaks are close in energy, below the energy resolution of our current data. This may theoretically happen when one of the following conditions is met: (i) $\tilde{U} \sim 0$; (ii) $\tilde{J} \sim 0$; or (iii) $\tilde{U} \sim \tilde{J}$ [Eq. (A4)], which also represents a strong constraint on the sign-changing scenario.

We propose that further STM experiment with higher energy resolution and a strong magnetic field will be helpful to clarify this issue. The external magnetic field effectively changes $\tilde{J}$. For the sign-changing scenario, it should split the resonance peak as long as $\tilde{U} \neq 0$. For the sign-preserving scenario, it shifts the peak without splitting.

**Acknowledgement**
This work is supported by NSFC (Grants No. 91421312 and No. 11574174) and MOST of China (Grant No. 2015CB921000). J.-H. M and Z. L acknowledge support from Tsinghua University Initiative Scientific Research Program and NSFC under Grant No. 11774196.

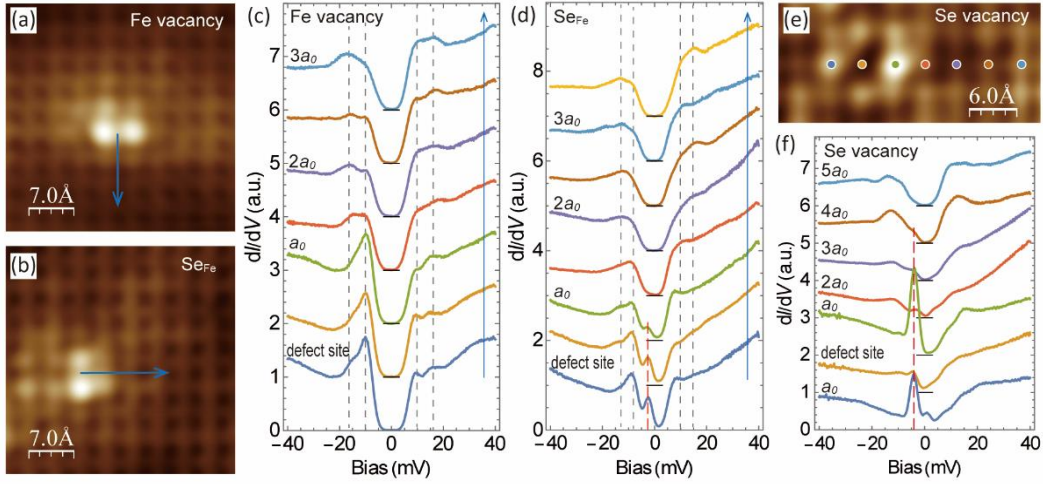

FIG. 1. (a) (b) and (e) Topographic images of monolayer FeSe on SrTiO$_3$(001) with native defects, such as Fe vacancy ( (a),$V$ = -100 mV, $I$ = 200 pA), Se$_{Fe}$ antisite defect ((b), $V$ = -80 mV, $I$ = 200 pA) and Se vacancy ((e), $V$ = -100 mV, $I$ = 200 pA). (c) and (d) Tunneling spectra taken along the blue lines in (a) and (b), respectively, with steps of 0.5 $a_0$. The grey dashed lines indicate the double coherence peaks in the energy gaps. (f) Tunneling spectra taken around Se vacancy as marked by the colored dots in (e). The red dashed lines in (d) and (f) label the peak positions of in-gap bound states. For (c), (d) and (f), $V$ = 40 mV, $I$ = 500 pA.

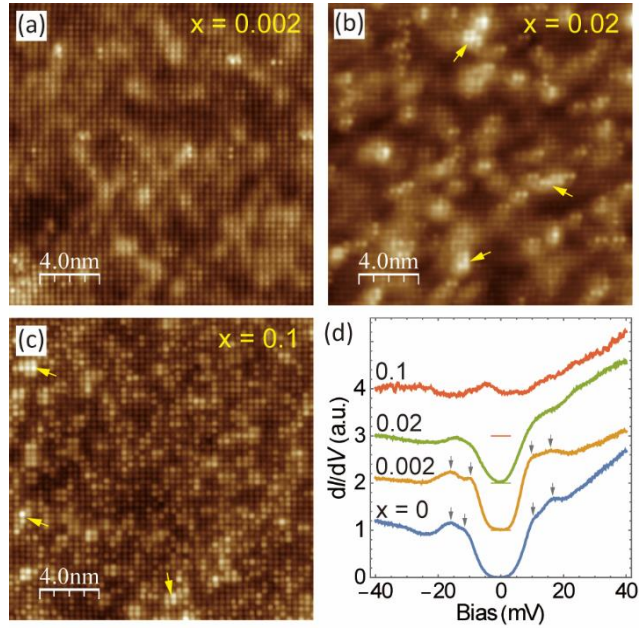

FIG. 2. (a)-(c) Topographic images of monolayer $Fe_{1-x}Co_xSe$ on $SrTiO_3(001)$ with x= 0.002, 0.02, and 0.1, respectively. (a) $V$ = 100 mV, $I$ = 200 pA. (b) $V$ = 100 mV, $I$ = 200 pA. (c) $V$ = 100 mV, $I$ = 100 pA. (d) Tunneling spectra taken at locations away from Co defects on monolayer $Fe_{1-x}Co_xSe$, with a *dI/dV* curve of pure monolayer FeSe as reference. The arrows indicate the typical double-peak features in the energy gaps ($V$ = 40 mV, $I$ = 500 pA for x = 0, 0.002 and 0.02. $V$ = 40 mV, $I$ = 200 pA for x = 0.1).

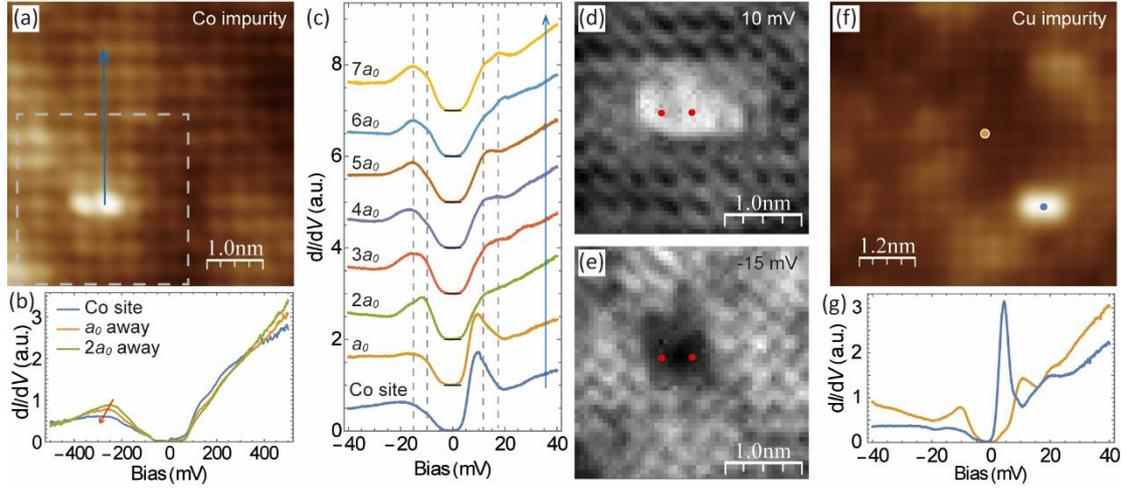

FIG. 3. (a) Topographic image of monolayer $Fe_{0.98}Co_{0.02}Se$ with single $Co_{Fe}$ substitution ($V$ = 70 mV, $I$ = 500 pA). (b) Tunneling spectra taken across $Co_{Fe}$ substitution ($V$ = 500 mV, $I$ = 500 pA, bias modulation amplitude = 5 mV) in a bias range from -500 mV to 500 mV. The arrow shows the energy shift of the kink. (c) Tunneling spectra taken across $Co_{Fe}$ substitution along the blue line in (a) with steps of $a_0$ ($V$ = 40 mV, $I$ = 500 pA) in a smaller bias range from -40 mV to 40 mV. The grey lines indicate the double coherence peaks in the energy gaps. (d) and (e) d$I$/d$V$ maps for the dashed square area in (a) at 10 mV and -15 mV, respectively (3 nm × 3 nm, $V$ = 40 mV, $I$ = 500 pA). The red dots mark the positions of the dumbbell-like pairs. (f) Topographic image of monolayer FeSe with single $Cu_{Fe}$ substitution ($V$ = 70 mV, $I$ = 500 pA). (g) Tunneling spectra taken on and far away from $Cu_{Fe}$ substitution as marked by the colored dots in (f) ($V$ = 40 mV, $I$ = 500 pA).

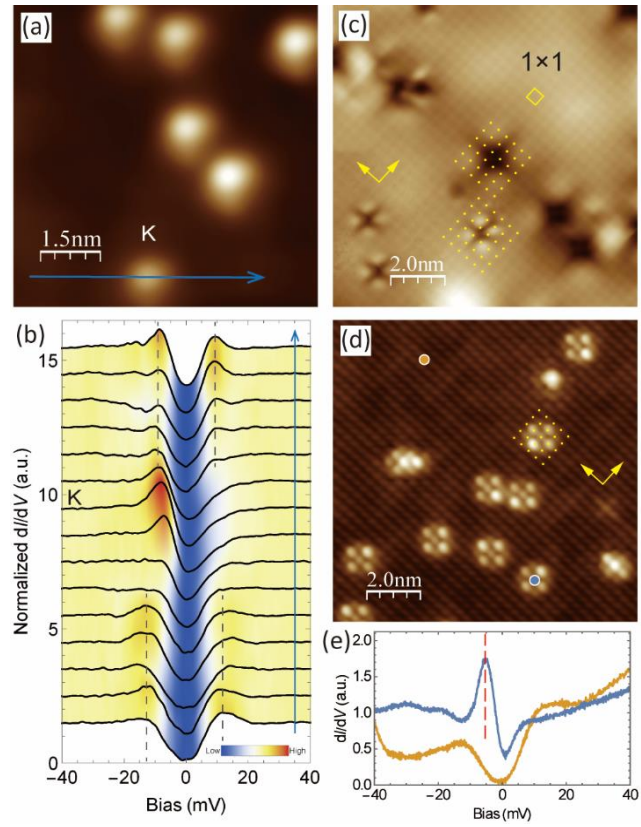

FIG. 4. (a) Topographic image of surface K-atoms adsorbed on monolayer FeSe ($V$ = 500 mV, $I$ = 50 pA). (b) Normalized tunneling spectra taken across a K-atom along the blue lines in (a) ($V$ = 50 mV, $I$ = 200 pA). The normalization is performed by dividing each raw d$I$/d$V$ curve by its background extracted from a polynomial fitting for $|V| > 20$ mV. (c) Topographic image of Sr-treated SrTiO$_3$(001) substrate ($V$ = 300 mV, $I$ = 200 pA). The yellow dots correspond to the (1 × 1) lattice. (d) Topographic image of monolayer FeSe grown on Sr-treated SrTiO$_3$(001) ($V$ = 50 mV, $I$ = 100 pA). The arrows in (c) and (d) indicate the in-plane lattice orientation. (e) Tunneling spectra taken on and far away from the Sr-related impurity as marked by the colored dots in (d) ($V$ = 50 mV, $I$ = 100 pA).

**Appendix: Derivation of the bound-state solution under the isotropic approximation**

In theory, a point impurity is commonly modeled as a delta potential, which induces scattering between different Bloch states:

$$\hat{V} = \sum_{k,k',\sigma=\pm} (U + J\sigma) c^+_{k\sigma} c_{k'\sigma}. \quad (A1)$$

In Eq. (1), U and J describe the strength of potential and magnetic scattering, respectively. The momentum dependence is neglected within the isotropic approximation. The impurity eigenvalue problem has been extensively studied by first including Eq. (A1) into the BCS mean-field Hamiltonian:

$$H_0 = \sum_{k,\sigma=\pm} \xi(k) c^+_{k\sigma} c_{k\sigma} - \sum_k \Delta(k) c^+_{k\sigma} c^+_{-k,-\sigma} + \text{h.c.,} \quad (A2)$$

where $\xi(k)$ is the band energy relative to the Fermi level and $\Delta(k)$ is the pairing order parameter. Then, diagonalize the total Hamiltonian $H = H_0 + \hat{V}$ by performing the Bogoliubov transformation:

$$\alpha_{k\uparrow} = u_{k\uparrow} c_{k\uparrow} - v_{k\downarrow} c^+_{-k\downarrow},$$
$$\alpha^+_{-k\downarrow} = u_{k\uparrow} c^+_{-k\downarrow} + v_{k\downarrow} c_{k\uparrow},$$

where $u^2_{k\uparrow} + v^2_{k\downarrow} = 1$.

Within the isotropic approximation, we allow sign change of $\Delta$ on different enclosed fermi surfaces, but treat its absolute value as a constant. This simple model is expected to capture the qualitative features of different pairing symmetry discussed in literature Ref. 38 (Fig. A1). Since it can be analytically solved, it also provides physically transparent pictures to examine against the experimental data.

(a) For the s-wave pairing case, we assume $\Delta(k) = \Delta$. Equations for the Bogoliubov coefficients are:

$$\begin{cases} -\varepsilon\ u_{k\uparrow} = -\xi_k u_{k\uparrow} - (U+J) u_{0\uparrow} - \Delta\ v_{k\downarrow} \\ \varepsilon\ v_{k\downarrow} = -\xi_k v_{k\downarrow} - (U-J) v_{0\downarrow} + \Delta\ u_{k\uparrow} \end{cases},$$

where $u_{0\uparrow} = \int d^2k\ u_{k,\uparrow} \equiv u_\uparrow(r=0)$, $v_{0\downarrow} = \int d^2k\ v_{k,\downarrow} \equiv v_\downarrow(r=0)$.

From these equations, we can write out the form of $u_{k\uparrow}$ ($v_{k\downarrow}$) in terms of $u_{0\uparrow}$ and $v_{0\downarrow}$. Then, by integrating over $k$ on the two sides, we get:

$$\begin{cases} \left[(U+J)\left(-\dfrac{\pi\ N_0\ \varepsilon}{\sqrt{\Delta^2 - \varepsilon^2}}\right) - 1\right] u_{0\uparrow} - (U-J)\left(-\dfrac{\pi\ N_0 \Delta}{\sqrt{\Delta^2 - \varepsilon^2}}\right) v_{0\downarrow} = 0 \\ (U+J)\left(-\dfrac{\pi\ N_0 \Delta}{\sqrt{\Delta^2 - \varepsilon^2}}\right) u_{0\uparrow} + \left[(U-J)\left(\dfrac{\pi\ N_0\ \varepsilon}{\sqrt{\Delta^2 - \varepsilon^2}}\right) - 1\right] v_{0\downarrow} = 0 \end{cases},$$

where $N_0$ is the density of states at the Fermi surface.

A bound state corresponds to nonzero $u_{0\uparrow}$ and $v_{0\downarrow}$. Such a solution exists only when

$$\begin{vmatrix} (U+J)\left(-\dfrac{\pi\ N_0\ \varepsilon}{\sqrt{\Delta^2 - \varepsilon^2}}\right) - 1 & -(U-J)\left(-\dfrac{\pi\ N_0 \Delta}{\sqrt{\Delta^2 - \varepsilon^2}}\right) \\ (U+J)\left(-\dfrac{\pi\ N_0 \Delta}{\sqrt{\Delta^2 - \varepsilon^2}}\right) & \left[(U-J)\left(\dfrac{\pi\ N_0\ \varepsilon}{\sqrt{\Delta^2 - \varepsilon^2}}\right) - 1\right] \end{vmatrix} = 0,$$

which determines the bound-state energy:

$$(\Delta_1 = \Delta_2 = \Delta): \qquad \varepsilon_{1,2} = \pm \Delta \frac{1 + (\tilde{U}^2 - \tilde{J}^2)}{\sqrt{\left[1 + (\tilde{U}^2 - \tilde{J}^2)\right]^2 + 4\tilde{J}^2}}, \quad (A3)$$

where $\tilde{U} = \pi N_0 U$ and $\tilde{J} = \pi N_0 J$ are the dimensionless scattering strength averaged on the FS. This solution is identical to Ref.39, as expected. It is evident that when $\tilde{J} = 0$, $\varepsilon = \pm \Delta$, independent of the value of $\widetilde{U}$, i.e. there is no well-defined bound state for nonmagnetic impurities. Note that the bound-state solution always come in pairs, and the positive-energy and negative-energy eigenstates are connected via a particle-hole transformation.

(b) For the sign-changing pairing case, we assume that the two Fermi surfaces (FS) have $\Delta_1 = -\Delta_2 = \Delta$. The Bogoliubov coefficients then also carry an index labeling the FS: $(u^1_{k\uparrow}, v^1_{k\downarrow}, u^2_{k\uparrow}, v^2_{k\downarrow})$. The two FS are coupled by the impurity scattering. The eigenvalue equations are:

$$\begin{cases} -\varepsilon\, u^1_{k\uparrow} = -\xi_k u^1_{k\uparrow} - (U+J)(u^1_{0\uparrow} + u^2_{0\uparrow}) - \Delta\, v^1_{k\downarrow} \\ \varepsilon\, v^1_{k\downarrow} = -\xi_k v^1_{k\downarrow} - (U-J)(v^1_{0\downarrow} + v^2_{0\downarrow}) + \Delta\, u^1_{k\uparrow} \\ -\varepsilon\, u^2_{k\uparrow} = -\xi_k u^2_{k\uparrow} - (U+J)(u^1_{0\uparrow} + u^2_{0\uparrow}) + \Delta\, v^2_{k\downarrow} \\ \varepsilon\, v^2_{k\downarrow} = -\xi_k v^2_{k\downarrow} - (U-J)(v^1_{0\downarrow} + v^2_{0\downarrow}) - \Delta\, u^2_{k\uparrow} \end{cases}$$

From these equations, we can write out the form of $u^{1,2}_{k\uparrow}$ ($v^{1,2}_{k\downarrow}$) in terms of $u^{1,2}_{0\uparrow}$ and $v^{1,2}_{0\downarrow}$. Then, by integrating over $k$ on the two sides, we get:

$$\begin{pmatrix} -1 - \frac{\pi N_1 (J+U)\varepsilon}{\sqrt{\Delta^2 - \varepsilon^2}} & \frac{\pi N_1 (-J+U)\Delta}{\sqrt{\Delta^2 - \varepsilon^2}} & -\frac{\pi N_1 (J+U)\varepsilon}{\sqrt{\Delta^2 - \varepsilon^2}} & \frac{\pi N_1 (-J+U)\Delta}{\sqrt{\Delta^2 - \varepsilon^2}} \\ -\frac{\pi N_1 (J+U)\Delta}{\sqrt{\Delta^2 - \varepsilon^2}} & -1 + \frac{\pi N_1 (-J+U)\varepsilon}{\sqrt{\Delta^2 - \varepsilon^2}} & -\frac{\pi N_1 (J+U)\Delta}{\sqrt{\Delta^2 - \varepsilon^2}} & \frac{\pi N_1 (-J+U)\varepsilon}{\sqrt{\Delta^2 - \varepsilon^2}} \\ -\frac{\pi N_2 (J+U)\varepsilon}{\sqrt{\Delta^2 - \varepsilon^2}} & -\frac{\pi N_2 (-J+U)\Delta}{\sqrt{\Delta^2 - \varepsilon^2}} & -1 - \frac{\pi N_2 (J+U)\varepsilon}{\sqrt{\Delta^2 - \varepsilon^2}} & -\frac{\pi N_2 (-J+U)\Delta}{\sqrt{\Delta^2 - \varepsilon^2}} \\ \frac{\pi N_2 (J+U)\Delta}{\sqrt{\Delta^2 - \varepsilon^2}} & \frac{\pi N_2 (-J+U)\varepsilon}{\sqrt{\Delta^2 - \varepsilon^2}} & \frac{\pi N_2 (J+U)\Delta}{\sqrt{\Delta^2 - \varepsilon^2}} & -1 + \frac{\pi N_2 (-J+U)\varepsilon}{\sqrt{\Delta^2 - \varepsilon^2}} \end{pmatrix} \begin{pmatrix} u^1_{0\uparrow} \\ v^1_{0\downarrow} \\ u^2_{0\uparrow} \\ v^2_{0\downarrow} \end{pmatrix} = 0,$$

where $N_{1,2} = N_0/2$ is the density of states associated with the two FS. The bound-state energy is determined by equating the determinant to zero:

$$(\Delta_1 = -\Delta_2 = \Delta): \quad \varepsilon_{1,2} = \pm \frac{\Delta}{\sqrt{1+(\widetilde{U}+\tilde{J})^2}}, \quad \varepsilon_{3,4} = \pm \frac{\Delta}{\sqrt{1+(\widetilde{U}-\tilde{J})^2}}. \quad (A4)$$

This result can be directly compared with the Eq. (18) in Ref. 31 for the magnetic-only scattering case and Eq. (3) in Ref. 33 for the nonmagnetic scattering case. These two previous works studied the impurity state in superconducting iron pnictides starting from a two-band Hamiltonian and a $s_\pm$ pairing symmetry. Equation (4) can be shown to be identical to the previous ones by converting the electron-hole Fermi surface to two sheets of electron Fermi surfaces and making the isotropic approximation.

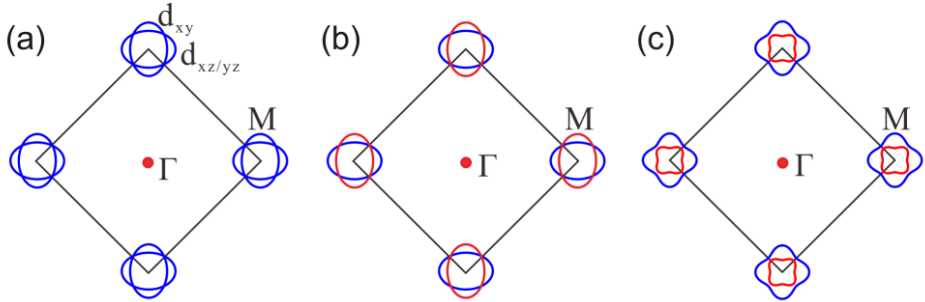

FIG. A1. Schematic representations of sign-preserving (a) and sign-changing (b) and (c) for $\Delta(k)$ in monolayer FeSe on SrTiO$_3$. Here, the color designates the sign of the gap function (blue: + and red: -). (Adapted from D. H. Lee, Chin. Phys. B, 24, 117405 [38])